\documentclass[11pt]{article}

\usepackage[utf8]{inputenc}
\usepackage[T1]{fontenc}
\usepackage{graphicx}
\usepackage{indentfirst}
\usepackage{authblk}
\usepackage{amsmath}
\usepackage{amssymb}
\usepackage{graphicx,color}
\usepackage{geometry}
\usepackage{csquotes}
\usepackage{cite}
\usepackage{braket}
\usepackage{physics}
\usepackage{subfigure}
\usepackage{xcolor}
\usepackage[title]{appendix}
\usepackage[normalem]{ulem}
\usepackage[colorlinks=true, linkcolor=blue, citecolor=blue, urlcolor=blue,linktoc=page]{hyperref} 

\title{\textbf{High-order coupling as a driver for Mott insulating behavior in Holography}}
\author[1,2]{Lin-Yue Bai\thanks{bailinyue137@163.com}}
\author[4]{René Meyer\thanks{rene.meyer@uni-wuerzburg.de}}
\author[3]{ Zhen-Hua Zhou\thanks{Corresponding author: dtplanck@163.com}}
\affil[1]{\small\itshape School of Physics and Astronomy, Beijing Normal University, Beijing 100875, China }
\affil[2]{\small\itshape Key Laboratory of Multiscale Spin Physics, Ministry of Education, Beijing Normal University, Beijing 100875, China}
\affil[3]{\small\itshape Institute for Theoretical Physics and Astrophysics\\ and\\
Würzburg-Dresden Cluster of Excellence ct.qmat\\
Julius-Maximilians-Universität Würzburg, 97074 Würzburg, Germany}
\affil[4]{\small\itshape School of Physics and Electronic Information, Yunnan Normal University, Kunming 650500, China}

\date{}

\begin{document}
\maketitle

\begin{abstract}
We construct a simple holographic model incorporating higher-order coupling terms for electron self-interactions. It can exhibit typical behavior of a Mott insulator, including a metal-insulator transition and a decrease in DC conductivity with the increase of charge density. In the analysis of AC conductivity, a soft gap is generally observed. Notably, when the DC conductivity approaches zero, the AC conductivity reveals a multi-peak structure, which can be attributed to the Mott and charge-transfer gaps observed experimentally in transition metals. With the increase of DC conductivity, the multi-peak structure gradually reverts to soft-gap behavior or even metallic conductivity. The accuracy of the numerical result is guaranteed by $\sigma(\omega\to 0)=\sigma_{DC}$ and sum rules.
\end{abstract}
\newpage

	\section{Introduction}
\begin{figure}[htbp]
  \renewcommand{\thesubfigure}{\Alph{subfigure}} 
  \centering

  \subfigure[]{\label{chargeMott1}
  \includegraphics[width=0.62\textwidth]{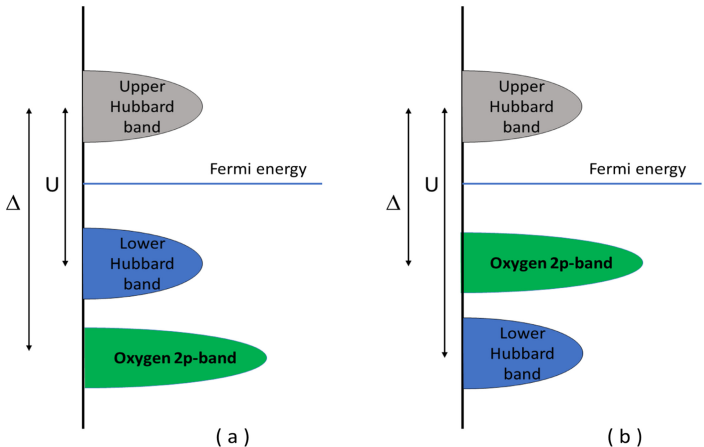}}\quad
  \subfigure[]{\label{ACCon1}
  \includegraphics[width=0.28\textwidth]{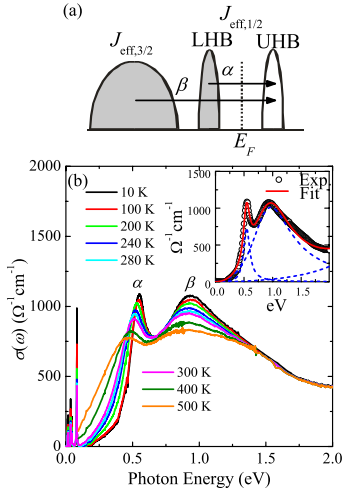}}
  \caption{Fig.\ref{chargeMott1}, adapted from \cite{roy2019mott}, show mechanism of Mott insulator, involving Hubbard band splitting and charge order, corresponds to the characteristic double-peak behavior observed experimentally of \ref{ACCon1}, reproduced from \cite{moon2009temperature}.}
  \label{condenseMott}
\end{figure}

The phenomenon of strongly correlated systems has attracted widespread research interest, with Mott insulators receiving significant attention as important systems for high-temperature superconductors \cite{lee2006doping}. In condensed matter physics, Mott insulators are typically described by the Hubbard model \cite{jeckelmann2000optical,fuhrmann2005mott,vzitko2015repulsive,patel2017non}, where the key mechanism is the dominance of local Coulomb interactions, leading to the formation of the Mott insulator state. The metal-insulator transition (MIT) \cite{chen2020metal,radonjic2010influence,ma2016correlated,duan2025breathing} in such systems arises from the competition between electronic kinetic energy and Coulomb self-interaction. The Coulomb interaction causes the splitting of the upper and lower Hubbard bands, thereby opening a new energy gap. Generally, this band splitting is accompanied by the emergence of charge order, which results in double-peak structure in the optical conductivity, corresponding to transitions within different bands and from the lower to the upper Hubbard band \cite{roy2019mott,moon2009temperature,park2014phonon,lovinger2020influence,wagner2023mott,yang2025gate,lunkenheimer2003dielectric,okamoto2004theory,gossling2008mott}, as shown in Fig.\ref{condenseMott}. This is a key signature for identifying the Mott insulator. Meanwhile, the divergence of DC resistivity requires vanishing conductivity at zero frequency, satisfying $\sigma(\omega=0)=0$ \cite{Andrade:2017ghg,carmelo2000finite}.
Gauge/gravity duality, as a commonly used approach in condensed matter physics \cite{Hartnoll:2009sz,Blake:2013bqa,Donos:2014cya,Donos:2014uba,Baggioli:2016oqk,Gulotta:2010cu,Ryu:2011vq,Grozdanov:2015qia,An:2020tkn,Blauvelt:2017koq,Cremonini:2018kla,Kiritsis:2016cpm,Charmousis:2010zz,Donos:2012ra,Andrade:2013gsa,Kim:2014bza}, has been employed to study Mott insulator. Investigations reveal that due to strong electron interactions, an effect analogous to "traffic jam" emerges with increasing charge density $q$, leading to a decrease in DC conductivity as the charge density $q$ rises. Meanwhile, DC conductivity exhibits the properties of metal-insulator transition (MIT) \cite{Baggioli:2016oju}, while the AC conductivity shows the behavior of soft gap.\cite{Baggioli:2016oju,Ling:2015epa,Kiritsis:2015oxa}.

In this work, we construct a holographic effective model incorporating strong interaction terms involving axion and gauge fields, aimed at simulating strong Coulomb interactions. Previous holographic models that have considered axion-gauge couplings include those in \cite{Megias:2016aje,Wu:2017mdl,Fu:2017oqa,Copetti:2017ywz,Li:2018vrz,Wu:2018pig,Cheng:2021hbw,Liu:2022bam,Zhong:2022mok,Li:2022yad,Liu:2022bdu,An:2020tkn,Baggioli:2016oqk,Bai:2023use,Gouteraux:2016wxj}. In our model, a metal-insulator transition (MIT) is consistently realized. At the same time, the DC conductivity exhibits a decrease with increasing charge density, described as a traffic jam effect and regarded as a typical characteristic of Mott insulator.
Furthermore, the AC conductivity of the model displays soft-gap behavior. It is noteworthy that as the DC conductivity approaches zero, the AC conductivity develops a multi-peak structure. It has been noted in \cite{wagner2023mott} that Mott insulator is expected to occur in the region where the Green's function equals to zero. The multi-peak structure associated with Mott insulator has been observed in experiments such as \cite{moon2009temperature,park2014phonon}. These two peaks can be interpreted as originating from charge-order-dominated level transitions and the splitting of the upper and lower Hubbard bands, respectively. In the numerical calculation of AC conductivity, the results from $\omega \to 0$ have a high consistency with DC conductivity, and the sum rules support the accuracy of the numerical method.

In section \ref{model}, we introduce the holographic model with high-order electromagnetic term and give the background solution. In section \ref{DC}, we derive the expression of DC conductivity, study its dependence on the charge density and temperature. A  "traffic jam" phenomenon and  the metal-insulator transition will be observed as  the characteristics of Mott insulators.  
In section \ref{AC}, we turn to  AC conductivity  to study the gap structure, including ubiquitous soft gap behavior and the emergence of multi-peak structures at vanishingly small DC conductivity. By comparing with experiments, we attribute these peaks to Mott and charge transfer gaps.

    \section{Holographic model}\label{model}
    
	We propose a 4-dimensional Einstein-Maxwell-axion model with gauge-axion coupling, for which the action is:
	\begin{align}\label{SXF4}
		\mathcal{S}=\int d^4 x \sqrt{-g}\left(R-2\Lambda-\frac{1}{4}F^2-V(X)-\mathcal{J}Tr[XF^{(4)}]\right)
		\end{align}
	where $R$ is Ricci scalar, $\Lambda=-3$ is negative cosmological constant and the $U(1)$ gauge field $F_{MN}\equiv\nabla_M A_N-\nabla_N A_M$.
    Here,  $A,B...$ represent bulk spacetime indices, $\mu,\nu...$ denote boundary indices, and $i,j...$ correspond to spatial indices.
	At the same time, momentum dissipation is incorporated through the introduction of the axion field $\phi^i=b x^i$. We define that
	\begin{align}
		&X_{MN}=\partial_M\phi^i\partial_N\phi^i\,,\quad\quad\quad X\equiv G^{MN}X_{MN}\,,\quad\quad\quad F_{AB}^{(2)}\equiv F_{AC}F_B^{~~C}\,,\nonumber\\
		& F_{AB}^{(4)}\equiv G^{CD}F_{AC}^{(2)}F_{BD}^{(2)}\,,\quad\quad\quad Tr[XF^{(4)}]\equiv X^{AB}F_{AB}^{(4)}\,.\nonumber
	\end{align}
    Throughout this paper, we choose $V(X)=\alpha X$ with the strength of the axion field denoted by $\alpha$ .
    We also introduce the gauge-axion coupling term $Tr[XF^{(4)}]$ where $\mathcal{J}$ represents the  the strength of coupling.
    The equations of motion can be derived as follow:
		\begin{subequations}
		\begin{align}
			&R_{AB}-(\frac{R}{2}+3)g_{AB}+\frac{1}{2}(\alpha X+\mathcal{J} Tr[XF^4])g_{AB}-\frac{1}{2}(F_{AC}F_{B}^{~~C}-\frac{1}{4}F^2 g_{AB})-\alpha \partial_A\phi^I\partial_B\phi^I\nonumber\\
			& \quad\quad\quad\quad\quad\quad -\mathcal{J}(XF^{(4)}+F^{(4)}X-FXFF^{(2)}-F^{(2)}FXF+F^{(2)}XF^{(2)})_{(AB)}=0\\
			&\nabla_A(F^{AB}-2\mathcal{J}(-XF^{(2)}F-FF^{(2)}X+FFXF+FXFF)^{AB})=0\\
			&\nabla_A(\alpha\partial^A\phi^i+\mathcal{J}(F^{(4)})^{AB}\partial_B\phi^i)=0
		\end{align}		
	\end{subequations}
        Those equations of motion admit asymptotically AdS dyonic black brane solutions that are given by
	\begin{subequations}
		\begin{align}
			&ds^2=\frac{1}{u^2}(-f(u)dt^2+\frac{1}{f(u)}du^2+dx^2+dy^2)\,,\qquad A=A_t(u)dt\,, \\
			&\phi^i=b x^i\,,\\
            &f(u)=1-\alpha b^2u^2-\frac{4-4\alpha b^2u_h^2+q^2u_h^4}{4u_h^3}u^3+\frac{q^2u^4}{4}\,,\label{fu}\\
			&A_t(u)=\mu-qu\,,
		\end{align}	
	\end{subequations}
	where $u_h$ denotes the horizon location. $\mu$ and $q$ are the chemical potential and the charge density, respectively.
	$b$ stands for the strength of the momentum dissipation. The temperature of the black brane is given by
	\begin{align}
		&T=-\frac{f'(u_h)}{4\pi}=\frac{3}{4 \pi u_h}-\frac{q^2 u_h^3}{16 \pi }-\frac{\alpha b^2 u_h}{4 \pi }\,,\label{T1} 
        \end{align}
Adding higher-order coupling terms does not alter the background solution. Therefore, thermodynamic quantities can be obtained through the standard renormalization result of \cite{Andrade:2013gsa}. The entropy density $s$, energy density  $\varepsilon$, and charge density $q$ can be expressed as:
     \begin{align}
         &s=\frac{4\pi}{u_h^2}\,,\quad\quad\quad\varepsilon=2(1-\alpha b^2 u_h^2+\frac{q^2u_h^4}{4})\,,\quad\quad\quad q=\frac{\mu}{u_h}\,.
     \end{align}

\section{ DC transport}\label{DC}
We calculate the DC conductivity following the standard process in \cite{Hartnoll:2009sz,Blake:2013bqa,Donos:2014cya,Donos:2014uba,Baggioli:2016oqk}. Turn on the following
consistent perturbations around the background:
\begin{align}
    &\delta A_x=-E_xt+a_x(u),\quad \delta g_{tx}=\frac{1}{u^2}h_{tx}(u),\quad \delta g_{ux}=\frac{1}{u^2}h_{ux}(u),\quad\delta \phi^{x}=\chi_x(u),
\end{align}
there exist an $u$-conserved quantity $\sqrt{-g}F^{ux}$ whose value  equals
to the electric current $J_x$ of the dual system at the boundary $u=0$. We can thus obtain its value at the
horizon $u=u_h$ in terms of the ingoing condition. 
Consequently, the expression for the DC conductivity $\sigma_{DC}$ can be derived by the Ohm’s law $J_x=E_x\sigma_{DC}$, which reads:
 \begin{align}\label{sigmax}
\sigma_{DC}=\frac{\left(1-2 \mathcal{J}b^2  q^2 u_h^6\right) \left(2 \alpha b^2+q^2 u_h^2\right)}{2 b^2 \left(\alpha+\mathcal{J} q^4 u_h^8\right)}\,.
 \end{align}
 
To achieve a zero conductivity, we require the coupling constant  $\mathcal{J}>0$. Then, we can define  $W\equiv 1-2\mathcal{J}b^2q^2u_h^6$ to analyze  the sign of $\sigma_{DC}$.
 Utilizing the expression of temperature \eqref{T1}, we find $W$ first decreases and then increases with increasing $q$, which satisfies: 
 \begin{subequations}\label{Wq}
 \begin{align}
 &\partial_q W(b,\alpha,\mathcal{J},T,q)<0\,, q\in[0,q_m)\,,\qquad\partial_q W(b,\alpha,\mathcal{J},T,q)>0\,, q\in(q_m,\infty)\,,\qquad\\
&q_m=\frac{2\sqrt{3-\alpha b^2 u_m^2-4\pi T u_m}}{u_m^2}\,,\qquad 
u_m=\frac{\sqrt{6\alpha b^2+9\pi^2T^2}-3\pi T}{2\alpha b^2}\,.
 \end{align}
 \end{subequations}
Therefore, the minimum value of $W$ is
\begin{align}
W_{min}(b,\alpha,\mathcal{J},T)=W(b,\alpha,\mathcal{J},T,q_m)=1-2\mathcal{J}b^2[q_m(\alpha,b,T)]^2[u_m(\alpha,b,T)]^6
\end{align}
It is easy to check $\partial_TW_{min}>0$, namely
\begin{align}
W(b,\alpha,\mathcal{J},T,q)>W_{min}(b,\alpha,\mathcal{J},T)>W_{min}(b,\alpha,\mathcal{J},0)=1-\frac{18\mathcal{J}}{\alpha}\,.
\end{align}
\begin{figure}[htbp]
  \renewcommand{\thesubfigure}{\alph{subfigure}} 
  \centering

  \subfigure[]{
  \includegraphics[width=0.31\textwidth]{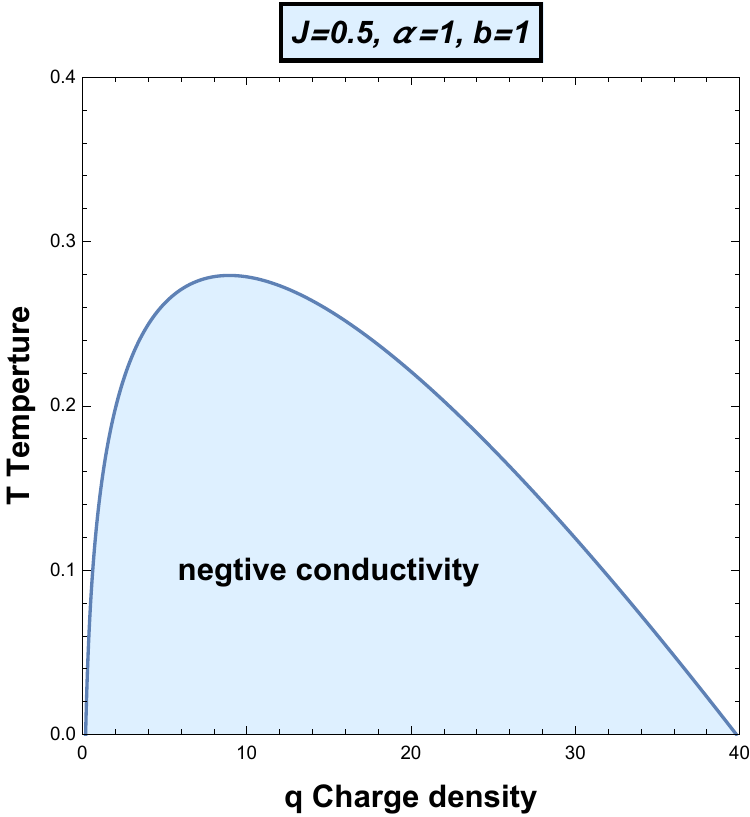} \label{xt1}}
  \subfigure[]{
  \includegraphics[width=0.31\textwidth]{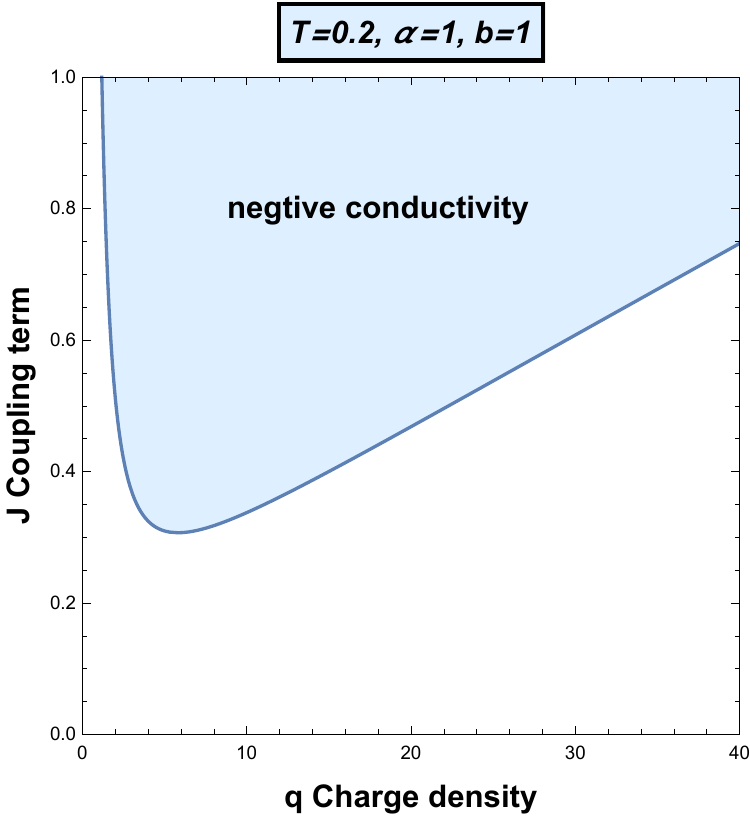} \label{xt2}}
  \subfigure[]{
  \includegraphics[width=0.31\textwidth]{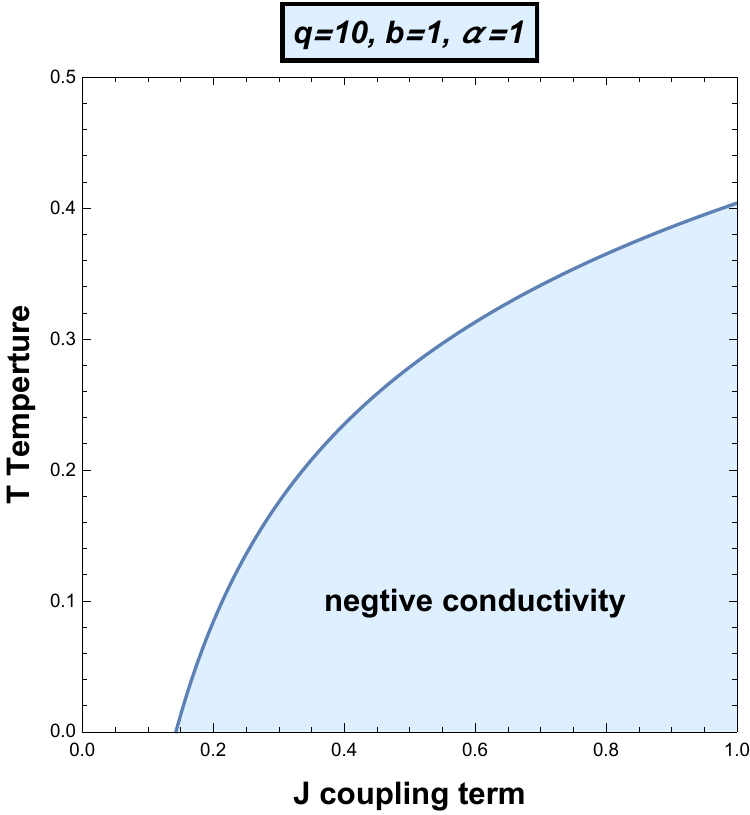} \label{xt3}}
  \caption{The negative conductivity region diagram when different parameters are fixed.}
  \label{xt}
\end{figure}
Thus, the DC conductivity is positive for all $T$ and $q$ when $\mathcal{J}<\alpha/18$. 
If $\mathcal{J}>\alpha/18$, there must exist $T_m$ satisfying $W_{min}(T_m)=0$, under which the DC conductivity can be negative.
According to \eqref{Wq}, there are two roots $q_1(T)$ and $q_2(T)$ of the equation $W(T,q)=0$ for a fixed $T<T_m$. Then, the region of negative conductivity is surrounded by the roots, as illustrated by Fig.\ref{xt1}. Similarly, we can also
find the  region of negative conductivity with fixed $T$ or $q$, as shown in  Fig.\ref{xt2} and  Fig.\ref{xt3} respectively. In the subsequent analysis, we should restrict the parameter range with positive conductivity, which can  also avoid ghosts and instabilities, (see  Appendix  \ref{sec:appendix}).

\begin{figure}[htbp]
  \renewcommand{\thesubfigure}{\alph{subfigure}} 
  \centering

  \subfigure[]{
  \includegraphics[width=0.45\textwidth]{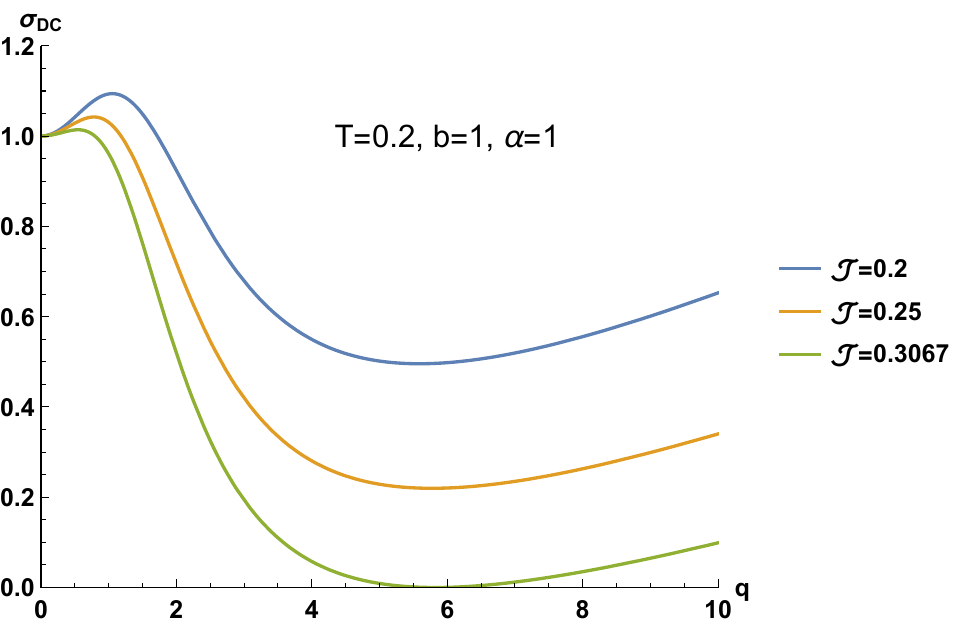} \label{DCqVJ}}
  \subfigure[]{
  \includegraphics[width=0.45\textwidth]{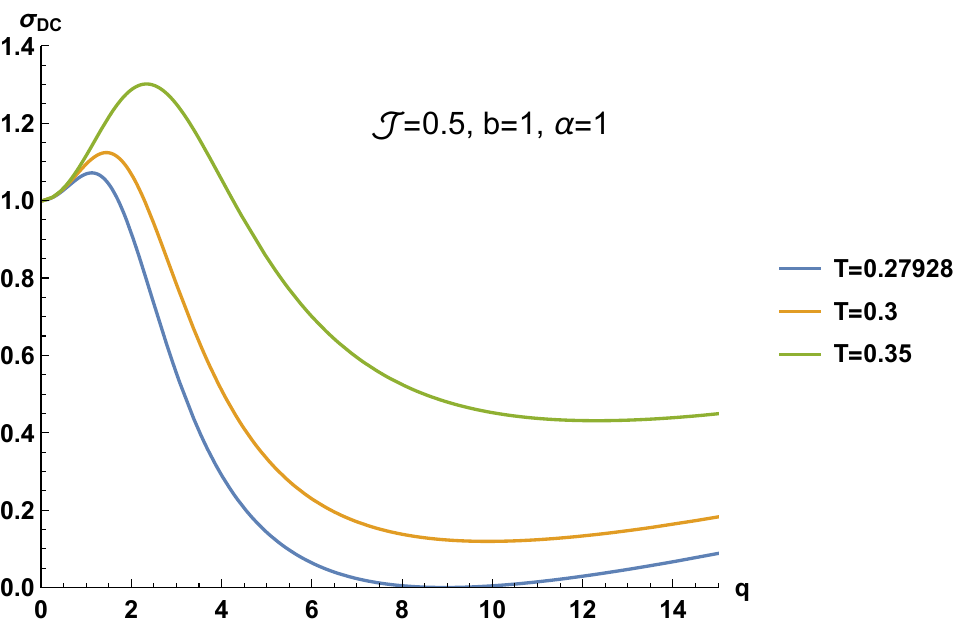} \label{DCqVT}}
  \caption{The DC conductivity varies with charge density $q$. It is observed that both an increase in coupling strength $\mathcal{J}$ and a decrease in temperature $T$ enhance the effect of conductivity reduction as charge density $q$ increases.}
  \label{DCq}
\end{figure}

Now we analyze the properties of DC conductivity. Firstly, we examine its variation with charge density. As shown in Fig.\ref{DCq}, $\sigma_{DC}(q)$ exhibits a non-monotonic variation pattern: as the charge density increases, the conductivity first increases, then decreases, and finally increases again.
\begin{figure}[htbp]
  \centering
  \includegraphics[width=0.6\textwidth]{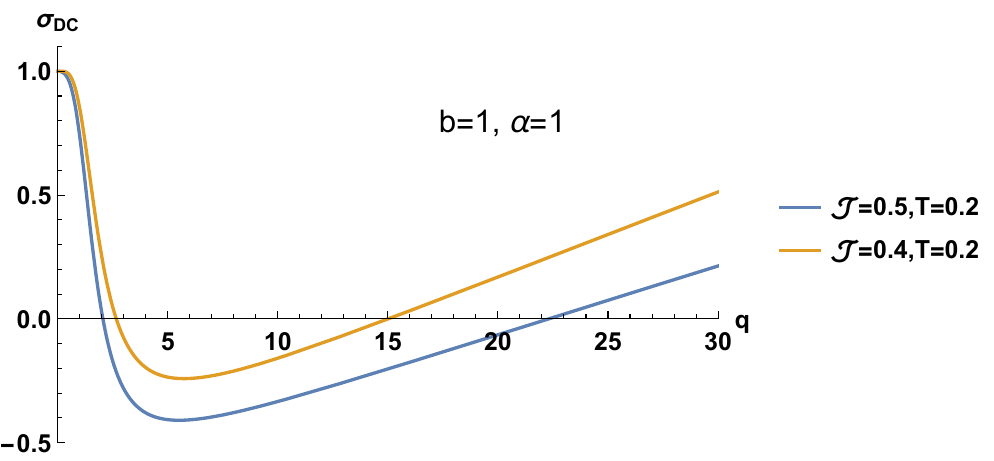} 
  \caption{The observation of negative DC conductivity suggests the opening of an energy gap.}
  \label{DCneg}
\end{figure}
The behavior where increasing charge density can suppress conductivity has been observed in previous holographic models with higher-order electromagnetic interactions, and is interpreted as the characteristic of Mott insulators caused by the "charge traffic jam" \cite{Baggioli:2016oju}. 

The overall non-monotonic behavior, on the other hand, should be attributed to the explanation of band theory and electron localization mechanisms. In the initial stage, the increase in conductivity corresponds to the  filling process of the conduction band. However, a continuing  increase of charge density must cause charge localization due to electron interactions, thereby  suppressing conductivity. If the charge density keeps increasing, it will eventually enter the process of filling a new empty band, resulting in the recovery of conductivity.
An increase in the coupling strength $\mathcal{J}$ enhances electron interactions, and a decrease in temperature $T$ reduces electron kinetic energy; both effects enhance electron localization, thereby suppressing the conductivity in Fig.\ref{DCq}.
 When parameters are selected from the negative conductivity region in phase diagram Fig.\ref{xt}, the system's conductivity decreases to zero and then enters a negative value regime in Fig.\ref{DCneg}. This behavior indicates the opening of an energy gap \cite{Seo:2023bdy}, and the system exhibits physical properties analogous to a Mott insulator which will be discussed in more detail in the next section.

\begin{figure}[htbp]
  \centering
  \includegraphics[width=0.6\textwidth]{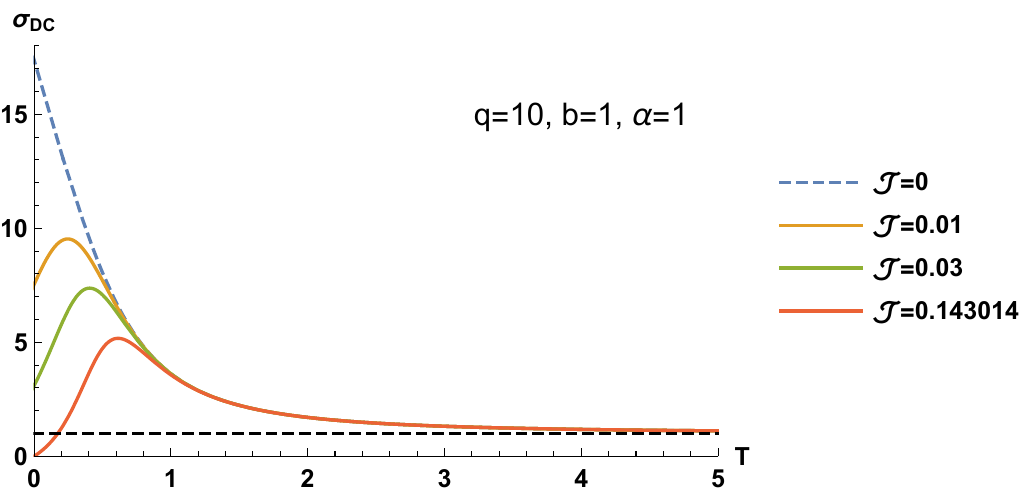} 
  \caption{The DC conductivity exhibits a metal-insulator transition as temperature changes.}
  \label{DCT}
\end{figure}
Finally, we examine the behavior of conductivity with temperature. As shown in Fig.\ref{DCT}, we can see $\partial\sigma_{DC}/\partial T>0$ signifies an insulating phase at lower temperature, and $\partial\sigma_{DC}/\partial T<0$ indicates a metallic phase at higher temperature\cite{Grozdanov:2015qia,Baggioli:2016oqk,An:2020tkn}. The condition $\sigma(T=0)=0$ (the red line in Fig.\ref{DCT}) further denotes the emergence of a good insulator \cite{An:2020tkn}.
It means the electron self-interaction can lead to   a metal-insulator transition (MIT) , which is also  an essential characteristic of a Mott insulator.

\section{AC conductivity}\label{AC}
As we see, the self-interaction term of electrons can suppress the DC conductivity, break its lower bound and make it approach zero or even become negative. This indicates the
existence of an energy gap, which we should further explore through the optical conductivity. We will find the self-interaction term generally leads to a soft-gap. Especially, as the DC conductivity approaches zero $\sigma_{DC}\rightarrow0$, the system develops multi-peak structure, which is the significant signature of Mott insulator \cite{okamoto2004theory,wagner2023mott,lunkenheimer2003dielectric,gossling2008mott,vzitko2015repulsive,Ling:2015epa}.

Explicitly, we turn on the following perturbations around the background:
\begin{align}
    \delta A_x=a_x(u,\omega)e^{-i\omega t},\qquad \delta\phi_x=\chi_x(u,\omega)e^{-i\omega t},\qquad \delta g_{tx}=h_{tx}(u,\omega)e^{-i\omega t},
\end{align}
the retarded Green function $G^R_{J^xJ^x}(\omega)$ of the electronic current
$J^x$ can be extracted by solving the perturbation equations \cite{Andrade:2013gsa,Kim:2014bza}. Then, the AC conductivity can be obtained by
\begin{align}
    \sigma_{xx}(\omega)\equiv\frac{1}{i\omega}G^R_{J^xJ^x}(\omega)\,.
\end{align}
The DC conductivity is defined as the value of the AC conductivity when the frequency $\omega$ approaches zero.
\begin{figure}[htbp]
  \centering
  \includegraphics[width=0.48\textwidth]{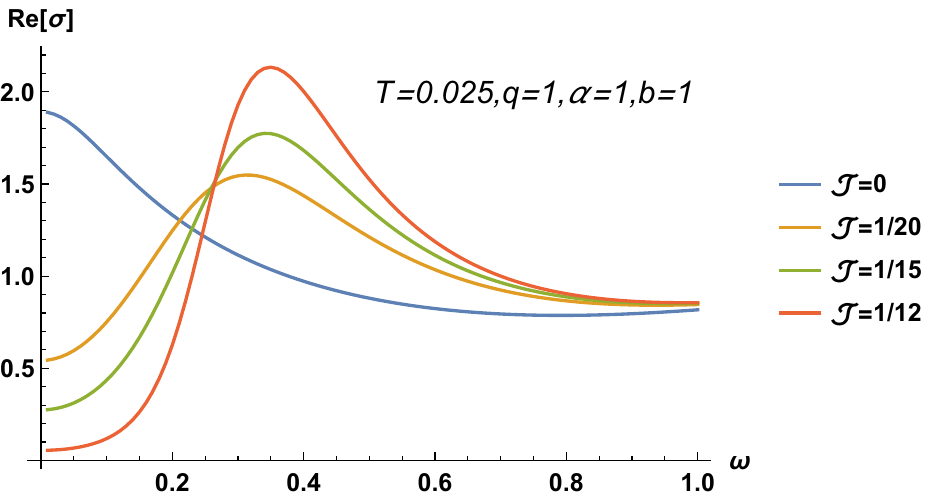} 
  \includegraphics[width=0.48\textwidth]{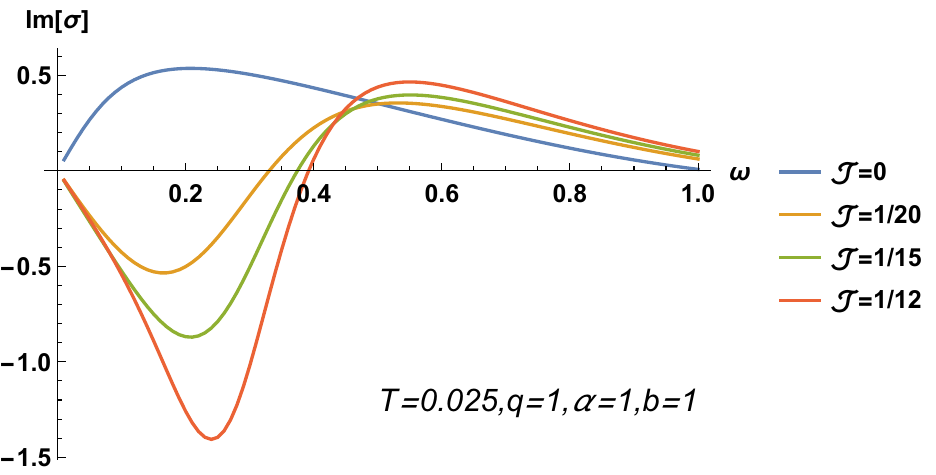} 
  \caption{At low temperatures, altering only the coupling parameter $\mathcal{J}$ leads to a more pronounced soft-gap peak.}
  \label{ACJ}
\end{figure}
\begin{figure}[htbp]
  \centering
  \includegraphics[width=0.48\textwidth]{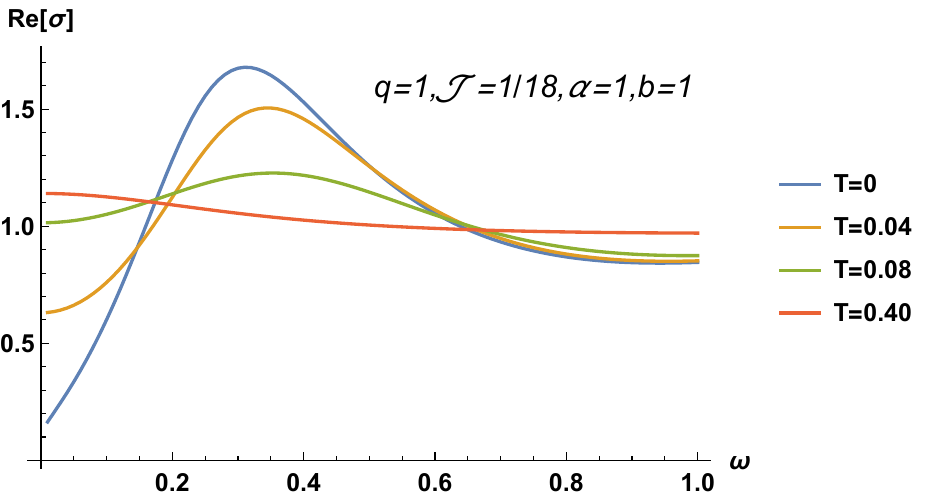} 
  \includegraphics[width=0.48\textwidth]{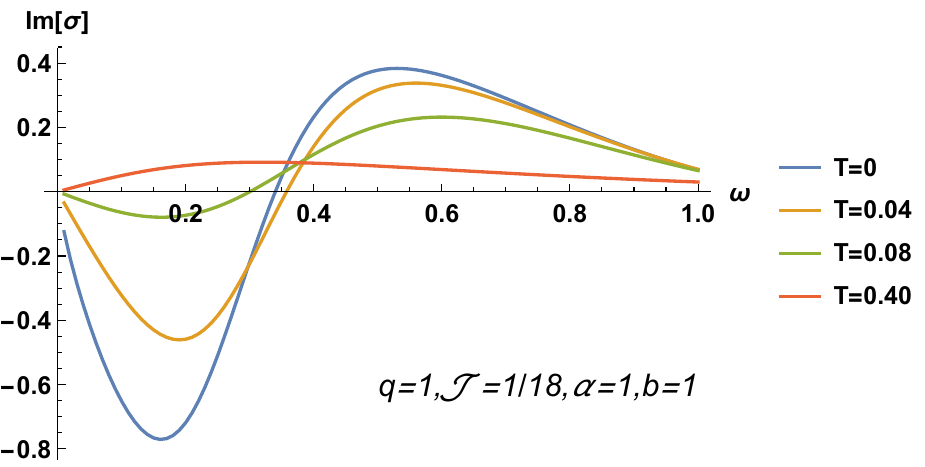} 
  \caption{A soft-gap emerges in our model. The soft-gap exhibits negligible temperature dependence, consistent with the anticipated temperature insensitivity characteristic of Mott insulator.}
  \label{ACT}
\end{figure}

We now present our numerical results. Firstly, we investigate the AC conductivity with different coupling strength $\mathcal{J}$, see Fig.\ref{ACJ}. The peak of $Re[\sigma(\omega)]$ represents the soft-gap, which becomes more pronounced with the increasing of coupling strength $\mathcal{J}$ and disappears at $\mathcal{J}=0$. Secondly, we can see from Fig.\ref{ACT}, as the temperature increases, the soft gap disappears and returns to the Drude peak of the simple axion model \cite{Andrade:2013gsa,Davison:2014lua}. This behavior can be understood within the framework of band theory: as thermal fluctuations excite electrons across the gap, the behavior of the soft gap gradually weakens with increasing temperature $T$. This can be understood within the framework of band theory, where  thermal fluctuations excite electrons across the gap.

\begin{figure}[ht]
  \centering
  \includegraphics[width=0.5\textwidth]{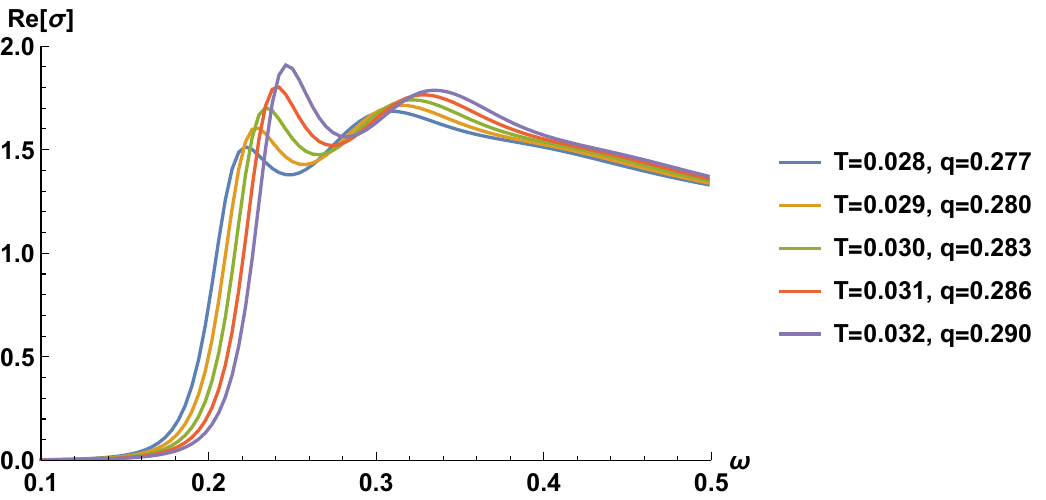}\\
  \caption{A typical  double-peak structure in AC conductivity, where $b=\alpha=1,\,\mathcal{J}=1/2$.}\label{EMXF4AC01}
\end{figure}
Notably, as the DC conductivity approaches zero, a distinct double-gap structure can be observed in the AC conductivity. 
As a representative case, we  fixed $b=\alpha=1,\,\mathcal{J}=1/2$. The temperature $T$ was varied from $0.028$ to $0.032$, while the corresponding charge density $q$ was adjusted  between  $0.277$ and $0.290$
to maintain the DC conductivity  around $10^{-3}$ \cite{Andrade:2017ghg,carmelo2000finite,vzitko2015repulsive}.The result of the sum rules is on the order of $10^{-4}$, calculated following the method in \cite{Ryu:2011vq,Gulotta:2010cu}. Consequently, the AC conductivity exhibited a characteristic double-peak structure, as illustrated in Fig.\ref{EMXF4AC01}. In comparison with experimental results for transition metals (Fig.\ref{ACCon1}), the left peak at lower excitation frequency can be attributed to the Mott gap
 arising from the splitting between the upper and lower Hubbard band. The right peak at higher excitation frequency
  may associate to the charge transfer gap between the inner orbital and the upper Hubbard band \cite{okamoto2004theory,wagner2023mott,lunkenheimer2003dielectric,gossling2008mott,vzitko2015repulsive}.
 \begin{figure}[htbp]
  \renewcommand{\thesubfigure}{\alph{subfigure}} 
  \centering
  \subfigure[]{
  \includegraphics[width=0.48\textwidth]{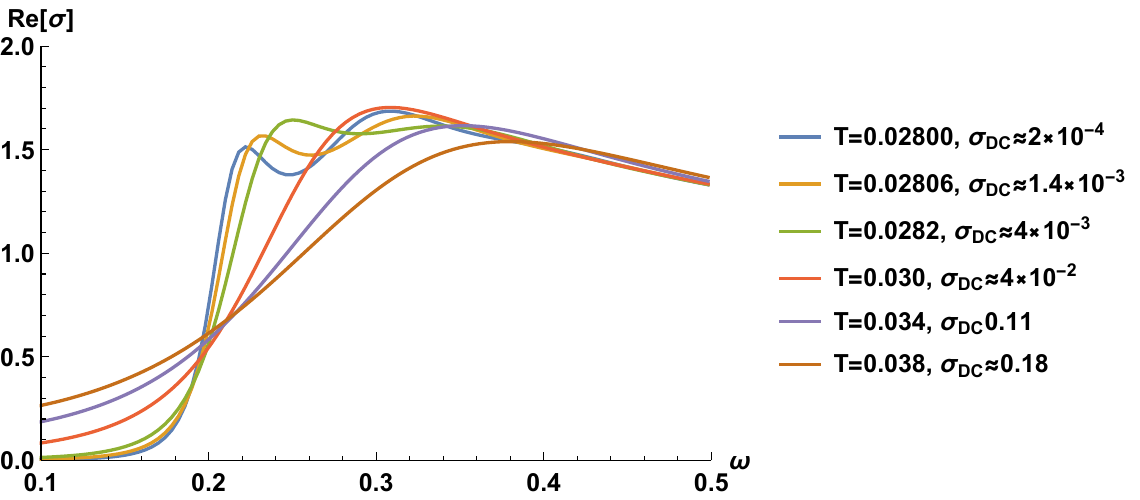} \label{EMXF4AC02a}}
  \subfigure[]{
  \includegraphics[width=0.47\textwidth]{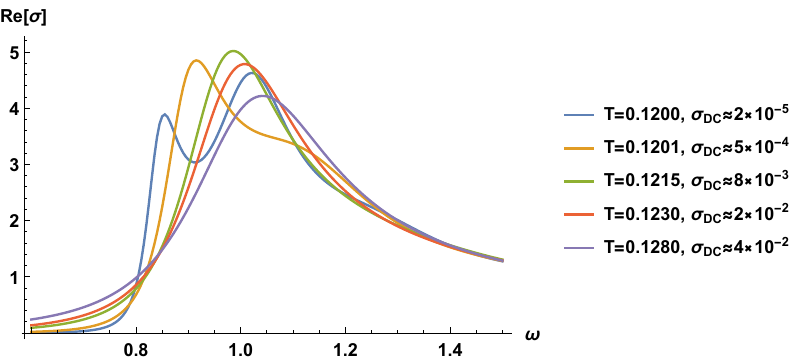} \label{EMXF4AC02b}}
  \caption{Double-peak becomes single-peak as the increase of temperature. Left: $b=\alpha=1,\,\mathcal{J}=1/2,\,q=0.277$. Right: $b=1,\alpha=5,\,\mathcal{J}=1/2,\,q=8.077$.}
  \label{EMXF4AC02}
\end{figure}

We find this double-peak structure is highly sensitive to parameter variations and appears only when the DC conductivity approaches a vanishingly small value.
As shown in figure \ref{EMXF4AC02}, the DC conductivity gradually rises  as the temperature increases. Concurrently, the  charge transfer gap (right-hand peak)  diminishes  and the Mott gap (left-hand peak) becomes increasingly prominent and  the initial double-peak   evolves into a single peak. With further increases in temperature, the Mott peak broadens and is suppressed, indicating a transition to a metallic state.

\begin{figure}[htbp]
  \renewcommand{\thesubfigure}{\alph{subfigure}} 
  \centering
  \subfigure[]{
  \includegraphics[width=0.47\textwidth]{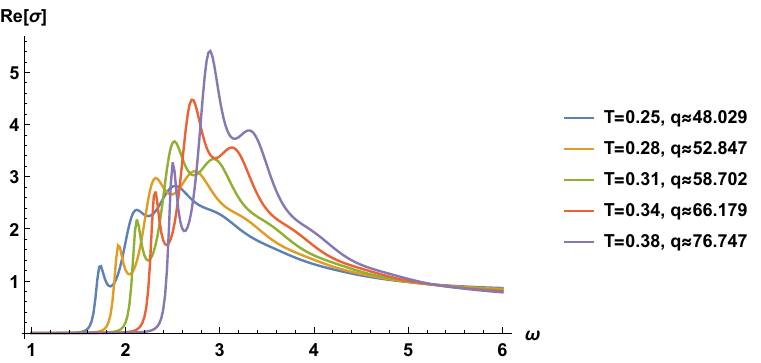} \label{EMXF4AC03a}}
  \subfigure[]{
  \includegraphics[width=0.48\textwidth]{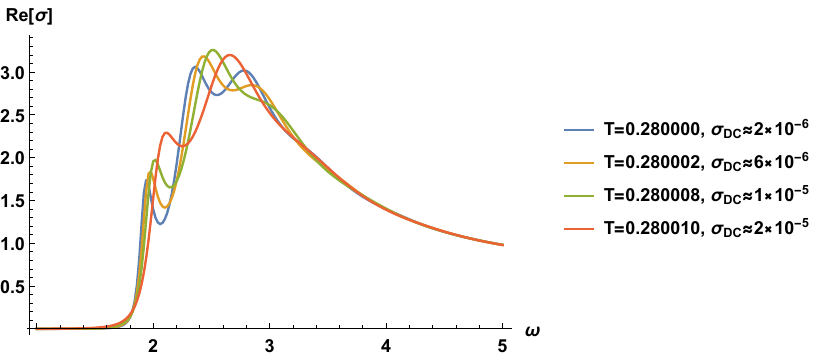} \label{EMXF4AC03b}}
  \caption{We set $b=3,\,\alpha=5,\,\mathcal{J}=1/2$. Left: typical three-peak structure. Right: three-peak becomes  double-peak and  single peak.}
  \label{EMXF4AC03}
\end{figure}

A vanishingly small value of the DC conductivity  is a characteristic feature of practical insulators. In such insulating phase, we expect to observe  richer gap structures in this system as reported in \cite{okazaki2018spectroscopic}. Indeed, as we reduce the DC conductivity  to about $ 10^{-6}$, a three-peak structure emerges  as illustrated in Fig.\ref{EMXF4AC03a}. An increase of the  temperature or DC conductivity will drive the three-peak structure into a double-peak (and eventually become a single peak), see Fig.\ref{EMXF4AC03b}. This multiplet structure originates from transitions between inter-orbital levels and the upper Hubbard band. As the system deviates from the insulating phase with zero DC conductivity, these discrete excitation peaks gradually broaden and merge, and their spectral weight is ultimately redistributed into the background conductivity of unity. These results suggest that the holographic model of charge self-interaction can capture the multi-gap structure observed in the insulating phase of transition metals at finite temperatures.

\section{Discussion}
Our holographic model reproduces Mott insulator-like behavior through higher-order coupling terms. In the study of DC conductivity, we observe a decrease in DC conductivity with increasing charge density q, which is manifested as "traffic jam". This phenomenon can be explained within the framework of band transport: enhanced electron filling increases the Coulomb interaction potential, forming bound states that restrict electron mobility. Further increase in charge density strengthens these bound states, thereby reducing conductivity.  The DC conductivity only begins to rise after the current-carrying band is completely filled and the next energy band starts to populate.

In the analysis of AC conductivity, a soft gap is generally observed. The Coulomb interaction enhances the soft-gap behavior, whereas increasing temperature weakens it. This is because the former promotes Mott-localization, while the latter enhances electron delocalization, leading to a reduction in the AC conductivity peak. When parameters are chosen such that the DC conductivity approaches zero, the AC conductivity develops a multi-peak structure, indicating the opening of new energy gaps. Analogous to the Hubbard model in condensed matter physics, when the upper Hubbard band splits into multiple sub-bands, the AC conductivity exhibits multiple peaks \cite{moon2009temperature,park2014phonon,lovinger2020influence,okazaki2018spectroscopic}. Our model demonstrates features reminiscent of Mott insulators, with the AC conductivity showing double or even triple-peak structures and the number of AC conductivity peaks increases as the DC conductivity decreases—a multi-peak signature observable in experiments. If a similar holographic model is constructed using a two-current setup, the AC conductivity can also exhibit a gap along with multi-peak characteristics. Work in this direction is currently underway.

\section*{Acknowledgments}
This work is supported by the National Natural Science Foundation of China under Grant No.12575047 and No.12361141825.

\begin{appendices}
\section{}\label{sec:appendix}  
\renewcommand{\thesection}{}  
\renewcommand{\thesubsection}{}  
In this appendix, we present the minimal consistency conditions to avoid ghosts and instabilities, following the approach outlined in \cite{Baggioli:2016oju,Baggioli:2016oqk}.

Firstly, we consider the transverse  vector and  scalar modes at the decoupling level by introducing the
electromagnetic  perturbation $\delta A_x=a_x(t,u,y)$ or the axional  perturbation $\delta\phi^x=\phi_T(t,u,y)$ separately. The corresponding perturbation equations are obtained as follows:
\begin{subequations}
\begin{align}
&\frac{\partial_u\big(f(u)(1-2 \mathcal{J} b^2 q^2 u^6)\partial_ua_x\big)+\partial^2_ya_x}{f(u)(1-2 \mathcal{J} b^2 q^2 u^6)}-\frac{\partial^2_ta_x}{f(u)^2}=\frac{J_x}{u^2 f(u) \left(1-2 \mathcal{J} b^2 q^2 u^6\right)},\label{Trax}\\
&\frac{u^2\partial_u\big(f(u) (\alpha/u^2 +\mathcal{J} q^4 u^6)\partial_u\phi_T\big)+\alpha \partial_y^2\phi_T}{f(u) \left(\alpha +\mathcal{J} q^4 u^8\right)}-\frac{\partial_t^2\phi_T}{f(u)^2}=\frac{J_\phi}{u^2f(u)(\mathcal{J} q^4 u^8+\alpha)}.\label{Trphix}
\end{align}
\end{subequations}
where we have introduced a ﬁctitious source $J^xa_x$ and $J_\phi\phi_T$ to keep
track of the normalization of the kinetic term. No ghosts condition requires:
\begin{align}
    (1-2 \mathcal{J} b^2 q^2 u^6)>0,\quad\quad\quad \mathcal{J} q^4 u^8+\alpha>0.
\end{align}
For $u=0$, the second inequality requires $\alpha>0$. For $u=u_h$, the above inequality 
gives $1-2 \mathcal{J} b^2 q^2 u_h^6>0$ as well as $\mathcal{J} q^4 u_h^8+\alpha>0$. Recall the
expression of the DC conductivity (\ref{sigmax}), 
we find no ghosts condition automatically requires that the conductivity is positive.
Conversely, if one choose  $\mathcal{J} >0$ and $1-2 \mathcal{J} b^2 q^2 u_h^6>0$ as is done in the main text, the
no ghosts condition is satisfied. 

Next, we consider the longitudinal vector and scalar modes. Turn on the following perturbations separately:
 \begin{subequations}
\begin{align}
&\delta A_\mu=(a_t(t,u,x,y),a_u(t,u,x,y),\partial_x\xi(t,u,x,y),\partial_y\xi(t,u,x,y)),\\
&\delta\phi^x=\phi_L(t,u,x,y),
\end{align}
 \end{subequations}
and fix $\delta A_t=a_t=0$ by gauge freedom,  we obtain the corresponding perturbation equations as follows:
\begin{subequations}
\begin{align}
& \nabla^2(a_u-\partial_u\xi)-\frac{\partial_t^2a_u}{f(u)(1-2 \mathcal{J}b^2  q^2 u^6)}=\frac{J_u}{u^2 \left(1-2 b^2 \mathcal{J} q^2 u^6\right)}\label{longAu}\\
&\frac{u^2\partial_u\big(f(u) (\alpha/u^2 +\mathcal{J} q^4 u^6)\partial_u\phi_L\big)+\alpha \nabla^2\phi_L}{f(u) \left(\alpha +\mathcal{J} q^4 u^8\right)}-\frac{\partial_t^2\phi_L}{f(u)^2}=\frac{J_\phi}{u^2f(u)(\mathcal{J} q^4 u^8+\alpha)}.\label{longphix}
\end{align}
\end{subequations}
The no ghosts condition is the same as the transverse case. 
Here, we can  read off  the ‘local’ speed of sound for longitudinal vector and scalar modes:
\begin{align}
 c_v^2=1-2 b^2 \mathcal{J} q^2 u^6\,,\qquad c_s^2=\frac{\alpha }{ \left(\alpha +\mathcal{J} q^4 u^8\right)}.
\end{align}
No gradient instabilities requires $c_v^2>0,\, c_s^2>0$, which is the same  as the no ghosts condition.
To summarize,  the positivity of the DC conductivity as well as $\mathcal{J} >0,\,\alpha>0$ can avoid ghosts and instabilities.

\end{appendices}

\newpage
\bibliographystyle{Ver}
\bibliography{ref}

\end{document}